\documentclass{JINST}
\usepackage{epsfig}

\title{Performance of ZnMoO$_4$ crystal as cryogenic scintillating bolometer to search for double 
beta decay of molybdenum}

\author{L.~Gironi$^{a,b}$, C.~Arnaboldi$^{a}$, J.~W.~Beeman$^{c}$, O.~Cremonesi$^{a}$, F.~A.~Danevich$^{d}$, V.~Ya.~Degoda$^{e}$,
L.~I.~Ivleva$^{f}$, L.~L.~Nagornaya$^{g}$, M.~Pavan$^{a,b}$, G.~Pessina $^{a}$, S.~Pirro$^{a}$\thanks{Corresponding author}~, 
V.~I.~Tretyak$^{d}$, I.~A.~Tupitsyna$^{g}$\\
\llap{$^a$} INFN - Sezione di Milano Bicocca,                           I 20126 Milano, Italy            \\
\llap{$^b$} Dipartimento di Fisica - Universit\`{a} di Milano Bicocca,  I 20126 Milano, Italy            \\ 
\llap{$^c$} Lawrence Berkeley National Laboratory,                      Berkeley, California 94720, USA  \\
\llap{$^d$} Institute for Nuclear Research,                             MSP 03680 Kyiv, Ukraine          \\
\llap{$^e$} Kyiv National Taras Shevchenko University,                  MSP 03680 Kyiv, Ukraine			 \\
\llap{$^f$} Institute for Scintillation Materials,                      61001 Kharkiv, Ukraine           \\
\llap{$^g$} Institute of General Physics,                               119991 Moscow, Russia            \\

E-mail: \email{Stefano.Pirro@mib.infn.it}}

\abstract{Zinc molybdate (ZnMoO$_4$) single crystals were grown for the first time by the Czochralski method and their luminescence was measured
under X ray excitation in the temperature range 85--400 K. Properties of ZnMoO$_4$ crystal as cryogenic low temperature
scintillator were checked for the first time. Radioactive contamination of the ZnMoO$_4$ crystal was estimated as $\leq0.3$ mBq/kg 
($^{228}$Th) and 8 mBq/kg ($^{226}$Ra). 
Thanks to the simultaneous  measurement of the scintillation light and the phonon signal, the 
$\alpha$ particles can be discriminated from the $\gamma$/$\beta$ interactions, making this compound  extremely promising 
for the search of neutrinoless Double Beta Decay of $^{100}$Mo. We also report on the ability to discriminate the $\alpha$-induced 
background without the light measurement, thanks to a different shape of the  thermal signal that characterizes $\gamma$/$\beta$ and $\alpha$ 
particle interactions.}

\begin{document}

\section{Introduction}

Neutrinoless Double Beta Decay ($0\nu2\beta$)  of atomic nuclei is a rare nuclear process able to give very important information about
properties of neutrino and weak interaction. 

Observations of neutrino oscillations \cite{Moh07,Fogli,Fogli-2} give a clear evidence that neutrino is a massive particle. While oscillation
experiments are sensitive to the neutrinos squared-mass differences, only the measurement of a $0\nu2\beta$ decay rate could establish the Majorana nature of the
neutrino, participate in the determination of the absolute scale of neutrino masses and test lepton number 
conservation  \cite{Vissani,reviewElliot,Avi08}.
Moreover, this process can clarify the presence of right-handed currents in weak interaction, and prove the existence of Majorons \cite{Avi08}. 
Taking into account ambiguity of theoretical estimations \cite{Sim-2008,Civ-2009,Men-2009,BandI-2009}, development of experimental methods for 
different $2\beta$ isotopes is highly requested.

$^{100}$Mo is one of the most promising $2\beta$ isotopes because of its large transition energy $Q_{2\beta}=3035$~keV~\cite{Aud03} and a considerable
natural isotopic abundance $\delta=9.67\%$~\cite{Boh05}. 
From the experimental point of view a large  $Q_{2\beta}$ value simplifies the  problem of background induced by natural radioactivity 
and cosmogenic activation. 

Nowadays the best sensitivity to $0\nu2\beta$ decay of $^{100}$Mo is the one reached by the NEMO experiment \cite{Arn05}
that, with $\simeq$7 kg of enriched $^{100}$Mo, has obtained a half-life limit $T_{1/2}^{0\nu}>4.6\times10^{23}$~yr at 90\% C.L.
Despite the beautiful result, the NEMO technique presents two disadvantages that limit the achievable sensitivity: the low detection 
efficiency of $0\nu2\beta$ events ($\approx14\%$)  and the poor energy resolution ($\approx10\%$ at the energy of
$Q_{2\beta}$ of $^{100}$Mo). Both these limitations can be overcome by the use of a high resolution detector containing in its sensitive 
volume the DBD candidate, i.e. by the use of the so called source=detector technique\footnote{For instance the detection efficiency
is $\approx86\%$ in the CUORICINO experiment with 0.75 kg tellurium oxide crystals \cite{Arna08} and $\approx93\%$ for Ge  
detectors \cite{Gomez2007}.}
In particular, the energy resolution (FWHM$~<1\%$) needed to investigate the normal hierarchy of the neutrino mass (half-life
sensitivity on the level of $10^{28}-10^{30}$ years) could be achieved only by bolometers or/and semiconductors providing an
energy resolution of about  few keV \cite{Zde04}. 

The main issue for the bolometric technique is the suppression -or active rejection- of the background induced by $\alpha$  emitters located 
close to the surface of a detector. Simulations show that this contribution will largely dominate the expected background of  the CUORE 
Experiment \cite{Arnaboldi2004,fondoBB} in the region of interest.   
The natural way to discriminate this background is to use a scintillating bolometer \cite{PHAN-2006}. 
In such a  device the simultaneous and independent read out of the heat and the scintillation light permits to discriminate events 
due to $\gamma$/$\beta$, $\alpha$ and neutrons  thanks to  their different scintillation yield.

Several inorganic scintillators containing molybdenum were developed in the last years. The most promising of them are molybdates of Calcium (CaMoO$_4$)
\cite{Belo05,Ann08}, Cadmium (CdMoO$_4$) \cite{PHAN-2006,Mikh06a}, Lead (PbMoO$_4$)
\cite{PHAN-2006,Dane10,Min92}, Lithium-Zinc (Li$_2$Zn$_2$(MoO$_4$)$_3$) \cite{Bash09}, and Lithium
(Li$_2$MoO$_4$) \cite{Bari09a,Bari09b}. However CaMoO$_4$ contains the $2\nu2\beta$ active isotope $^{48}$Ca which, even if present 
in natural Ca with a very small abundance of $\delta=0.187\%$ \cite{Boh05}, creates background at $Q_{2\beta}$ energy of $^{100}$Mo \cite{Ann08}.
CdMoO$_4$ contains the $\beta$ active $^{113}$Cd ($T_{1/2}=8.04\times 10^{15}$ yr \cite{Bell07}, $\delta=12.22\%$ \cite{Boh05}) which, 
besides being beta active, has a very high cross section to capture thermal neutrons. A potential disadvantage of PbMoO$_4$ (supposing crystals would be
produced from low-radioactive ancient lead \cite{Ales98-b,Dane09}) is that $^{100}$Mo would be only 27\% of the total mass. Finally
Li$_2$Zn$_2$(MoO$_{4})_{3}$ and Li$_2$MoO$_4$ have low Light Yield (LY).

ZnMoO$_4$ crystals were developed only very recently \cite{Ivle08,Nago08}. An important advantage of ZnMoO$_4$ is the absence of heavy elements
and high concentration of molybdenum (43\% in weight). 

The purpose of our work is to investigate  ZnMoO$_4$ crystals as  scintillating bolometers to search
for double beta decay of $^{100}$Mo. Luminescence of material was studied under X ray irradiation. The performances of the detector
were measured  through the use of the bolometric technique.

\section{Growth of ZnMoO$_4$ crystals}

The zinc molybdate (ZnMoO$_4$) charge was obtained by a solid-phase synthesis technique from MoO$_3$ and ZnO powders (both
of 99.995\% purity). Single crystals of up to dia=30 mm, h=40 mm were grown by the Czochralski technique with a drawing speed of 1.9
mm/h. The material is highly inert; the melting point is at ($1003\pm5$) $^\circ$C. The crystal density calculated from the X-ray data
is 4.317 g/cm$^3$, while the density measured by the pycnometric method is 4.19 g/cm$^3$ \cite{Ivle08}. Properties of ZnMoO$_4$
crystals are presented in Tab.~\ref{Tab:tab1}.

\begin{table}[tbp]
\caption{Properties of ZnMoO$_4$ crystal scintillators}
\label{Tab:tab1}
\begin{center}
\begin{tabular}{|l|l|l|}

 \hline

 Density (g/cm$^3$)                             &  4.3          & \cite{Ivle08}     \\

 Melting point ($^\circ$C)                      &  $1003\pm5$   & \cite{Ivle08}     \\

 Structural type                                &  Triclinic, $P1$ & \cite{Ivle08,Reic00} \\

 Cleavage plane                                 &  Weak (001)   & \cite{Ivle08}     \\

 Wavelength of emission maximum$^{\ast}$ (nm)   & 544           & Present work      \\

 \hline
 \multicolumn{3}{l}{$^{\ast}$Under X-ray excitation at room temperature.}           \\

 \end{tabular}
 \end{center}
 \end{table}

\section{Luminescence under X-ray excitation}

We measured the luminescence of ZnMoO$_4$ crystal  in the temperature interval 85--400 K under X-ray excitation. A sample of
ZnMoO$_4$ crystal ($10\times10\times2$~mm$^3$) was irradiated by X-ray from a BHV7 tube with a rhenium anode (20 kV, 20 mA). 
The light emitted by the crystal  was detected in the visible region by a FEU-106 photomultiplier (sensitive in the wide wavelength region of
$300-800$ nm). Spectral measurements were carried out using a high-aperture MDP-2 monochromator.

\begin{figure}[t]
\begin{center}
\mbox{\epsfig{figure=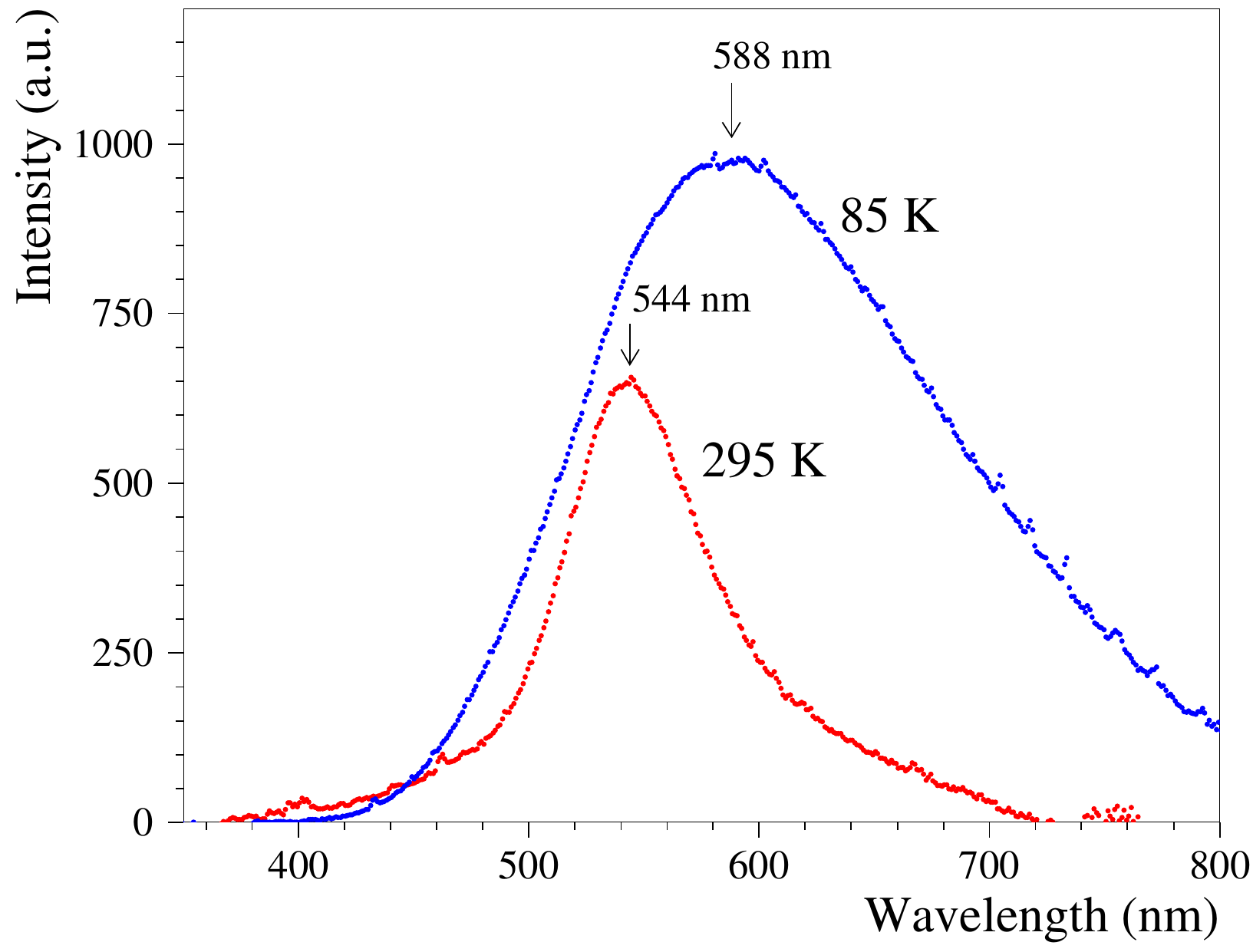,height=8.0cm}} \caption{(Color
online) Emission spectrum of ZnMoO$_4$ crystal under X-ray excitation at 85 K and 295 K.}
\label{fig:fig1}
\end{center}
\end{figure}

\begin{figure}[t]
\begin{center}
\mbox{\epsfig{figure=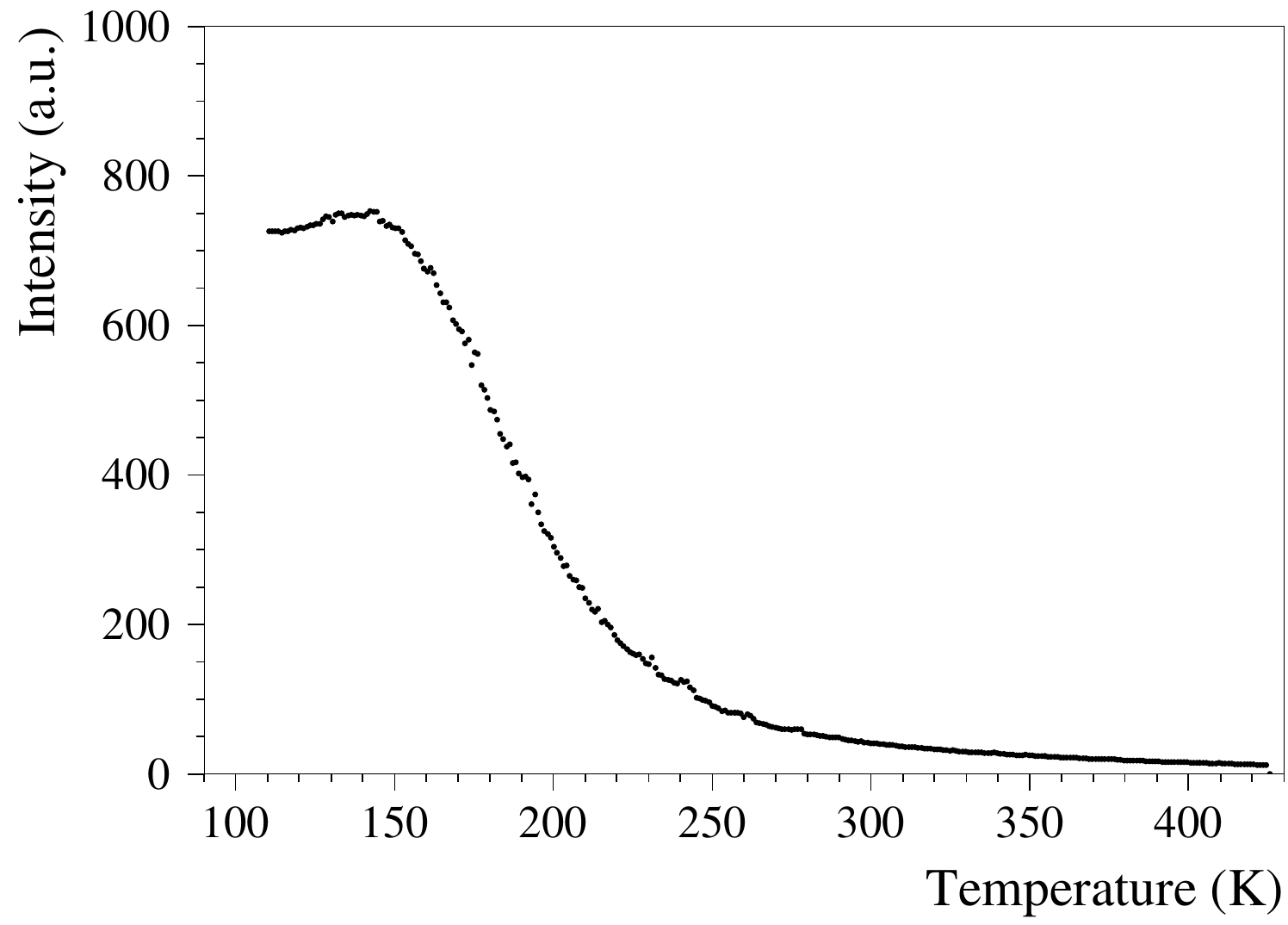,height=8.0cm}} \caption{Temperature
dependence of ZnMoO$_4$ luminescence intensity under X-ray
excitation.}
\label{fig:fig2}
\end{center}
\end{figure}

\begin{figure}[t]
\label{fig:fig3}
\begin{center}
\mbox{\epsfig{figure=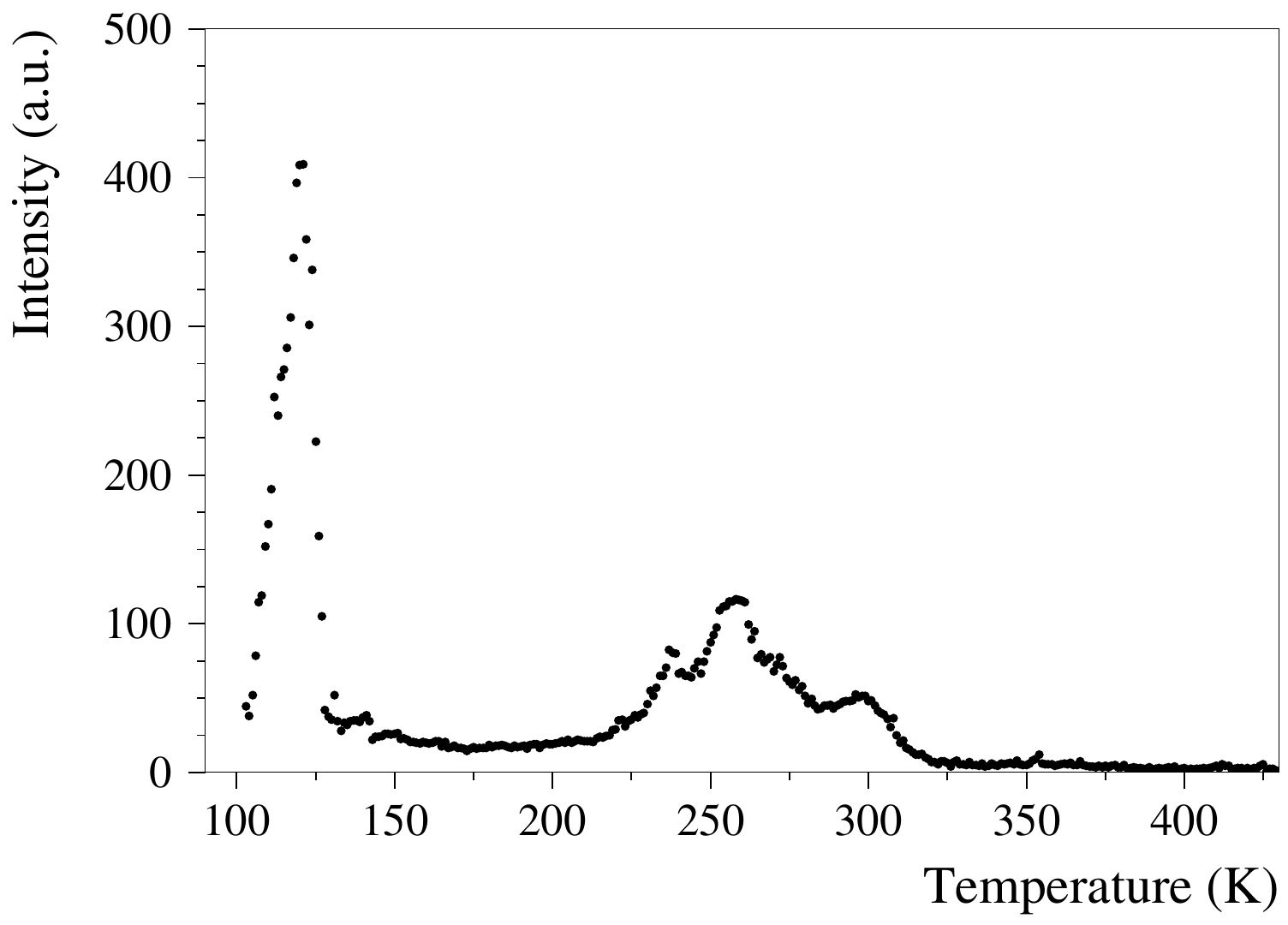,height=8.0cm}}
\caption{Thermoluminescence of ZnMoO$_4$ crystal after X-ray
excitation at liquid nitrogen temperature.}
\end{center}
\end{figure}

One band in the visible region with the maximum at 544 nm was observed at room temperature (Fig.~\ref{fig:fig1}). 
The radioluminescence (RL) spectrum,
measured at 85 K, is shifted to 588 nm. In addition, the spectrum measured at 85 K is much wider and distributed in the wavelength
region above 700 nm.
The temperature dependence of luminescence is presented in Fig.~\ref{fig:fig2}. The intensity of  luminescence slowly falls above 150 K,
but it still remains observable even at $\approx400$ K temperature.
A sharp peak of termostimulated luminescence (TSL) was observed at $\approx120$ K and  across the wide temperature range  
220--320 K (see Fig.~\ref{fig:fig3}),
with maxima at $\approx230$, $\approx260$, and $\approx300$ K. The TSL observed in our measurements indicates the 
presence of traps in the ZnMoO$_4$ sample. We assume it is due to defects and impurities in the
crystal.  Both can deteriorate the scintillation and/or the bolometric performances. Therefore, further R\&D is necessary to improve the 
quality of ZnMoO$_4$ crystals.

\section{ZnMoO$_4$ as a scintillating bolometer}

\begin{figure}
\begin{center}
\resizebox{0.7\textwidth}{!}
{\includegraphics{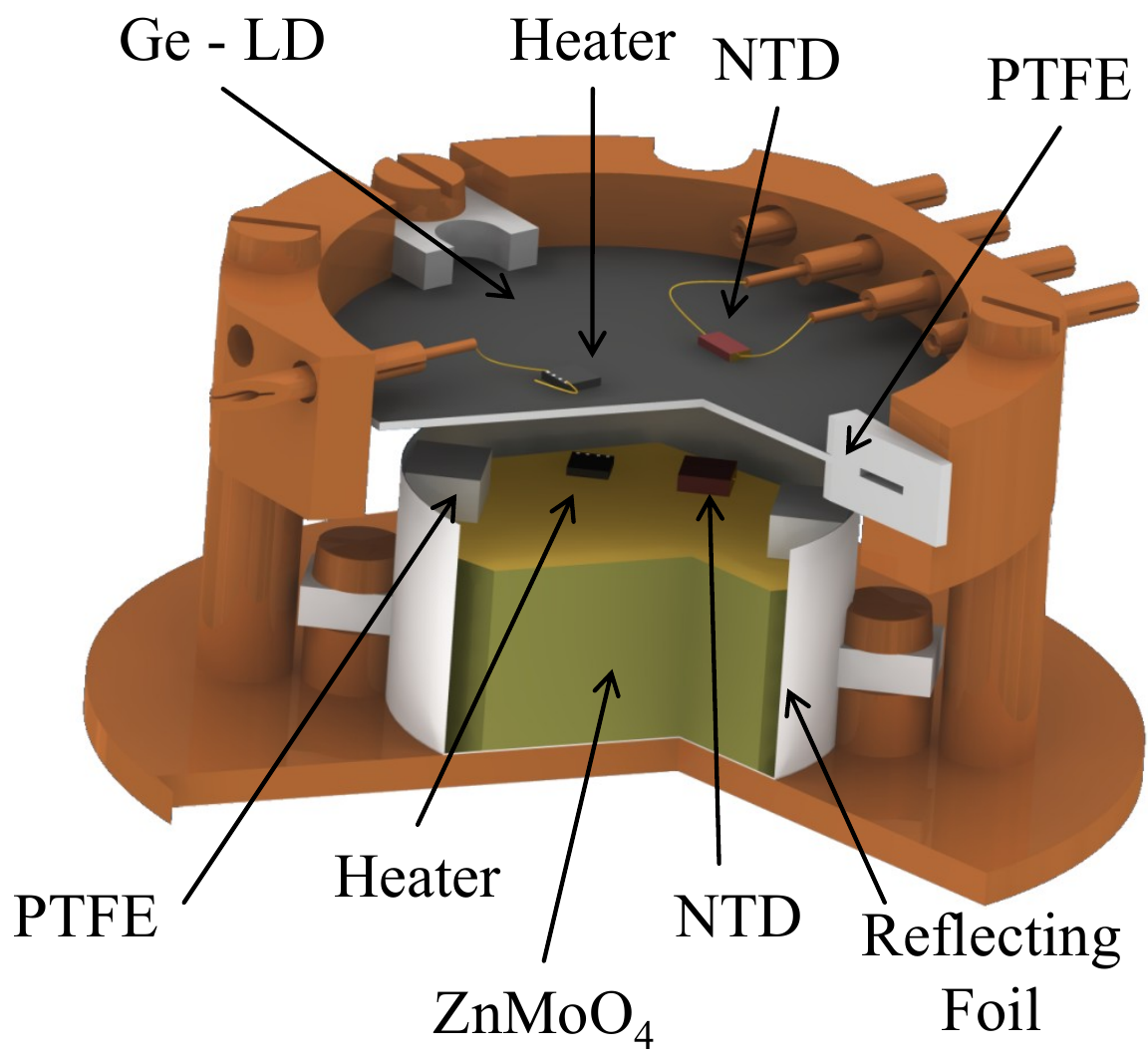}}
\caption{Setup of the detector.}
\label{fig:fig4}      
\end{center}
\end{figure}

We present in this section the results obtained with a small sample (19.8 g) of ZnMoO$_4$ operated as scintillating bolometer.
The shape of our sample is a regular hexagon, with a diagonal of 25 mm and a height of 11 mm. The color of the sample is fairly orange, showing evident inclusions 
along the central axis of the growth. 
The composite device (bolometer + light detector)   is schematized in Fig.~\ref{fig:fig4}. 
The crystal   is held by means of two L-shaped Teflon (PTFE) pieces fixed to  two cylindrical Cu frames; the PTFE   forces the crystal to the 
base consisting of a Cu plate covered with an Al foil. 
The crystal is surrounded by a  25.1 mm diameter  cylindrical reflecting foil (3M  VM2002).
At cryogenic temperatures (10$\div$100 mK) for which a detector can work as bolometer, no  ``standard'' light detectors can work properly.
The best way to overcome this problem is to use a second -very sensitive- ``dark'' bolometer that absorbs the scintillation light giving rise
to a measurable increase of its temperature. 
Our Light Detector (LD) \cite{NIMA-2006}  consists of  a  36 mm diameter, 1 mm thick pure Ge crystal absorber.

The temperature sensor of the ZnMoO$_4$ crystal  is a 3x3x1 mm$^3$ neutron transmutation doped Germanium thermistor, identical to the ones  used 
in the CUORICINO  experiment \cite{PRC-2008}.
The temperature sensor of the LD has a smaller volume (3x1.5x0.4 mm$^3$) in order to decrease its heat capacity, increasing therefore its 
thermal signal.
A resistor of  $\sim$300 k$\Omega$, realized with a heavily doped  meander on a 3.5 mm$^3$ silicon chip, is attached to  
each absorber and acts as a heater to  stabilize the gain of the bolometer \cite{ALES98,Arna2003}. 
The detectors were operated deep underground in the Gran Sasso National Laboratories in the CUORE R\&D test cryostat. 
The details of the electronics and the cryogenic facility can be found elsewhere \cite{NIMA-2006-B,NIMA-2006-C,NIMA-2004}.

The heat and light pulses, produced by a particle interacting in  the ZnMoO$_4$ crystal and transduced in a voltage pulse by 
the NTD thermistors, are amplified and fed into a 16 bit NI 6225 USB ADC unit. 
The entire waveform (\emph{raw pulse})  of each triggered voltage pulse is sampled and acquired.
The amplitude V$^{heat}$ and the shape of the voltage pulse is then determined by the off-line analysis that makes use of the 
Optimal Filter (OF) technique \cite{Arnaboldi-2004,GattiManfredi-86}. The signal amplitudes are computed as the maximum of the optimally 
filtered pulse, while the signal shape is evaluated on the basis of four different parameters: $\tau_{rise}$ and $\tau_{decay}$, TVL and TVR. 

$\tau_{rise}$ (the rise time) and $\tau_{decay}$ (the decay time) are evaluated on the \emph{raw pulse} as (t$_{90\%}$-t$_{10\%}$) and 
(t$_{30\%}$-t$_{90\%}$) respectively. The rise time is  dominated by the time constant of the absorber-glue-sensor interface  (as well 
as the heat capacity), while the decay time is determined by the crystal heat capacity and by its thermal conductance toward the heat sink. 

TVR (Test Value Right) and TVL (Test Value Left) are computed on the optimally filtered pulse as the least square differences with respect 
to the filtered 
response function\footnote{The response function of the detector, i.e. the shape of a pulse in absence of noise, is computed with a proper 
average over a large 
number of raw pulses. It is also used, together with the measured noise power spectrum, to construct the transfer function of the 
Optimal Filter.} of 
the detector: TVR on the right and TVL on the left side of the optimally filtered pulse maximum. These two parameters do not have a 
direct physical 
meaning, however they are extremely  sensitive (even in noisy conditions) to any difference between the shape of the analyzed pulse and the 
response function. Consequently, they are used either to reject fake triggered signals  or to identify variations in the pulse shape 
with respect to the response function (and this will be our case). \\

The thermal pulses are acquired within a 512 ms  time  window with a  sampling rate of 2~kHz. The trigger of the ZnMoO$_4$ is software generated while 
the LD is automatically acquired in coincidence with  the former. 

The energy calibration of the ZnMoO$_4$ crystal is performed  using   $\gamma$ sources ($^{232}$Th) placed outside the cryostat. 
The energy calibration of the LD, on the contrary, is obtained thanks to a  weak $^{55}$Fe source placed close to the Ge that illuminates 
homogeneously the face opposed to the  ZnMoO$_4$ crystal. During the LD calibration its trigger is set independent from 
the one of the ZnMoO$_4$.  The LD  is calibrated using a simple linear function. 
The main parameters of the two bolometers are listed in Tab.~\ref{Tab:tab2}.

\begin{table}[htb]
\begin{center}
\caption{Main parameters of the ZnMoO$_4$ bolometer  and light detector. The OF (Optimum Filter) FWHM represent the theoretical 
energy resolution as evaluated from the signal-to-noise ratio.The (absolute) Signal represents the voltage read across the thermistor for a unitary 
deposition of energy.}
\label{Tab:tab2}
\begin{tabular}{@{}lccccc}
\hline
             & Resistance     & OF FWHM   & $\tau_{rise}$   &  $\tau_{decay}$ 			& Signal         \\

             &  [M$\Omega$]   &  [keV] 	  &    [ms]		    &      [ms]	                & [$\mu$V/MeV]   \\

\hline
ZnMoO$_4$    &  2.5       	  &	   1.6      &     12.2          &       59.5                 &    107          \\
LD    	     &  14.5	              &   0.48     &     4.0           &       14.3                  &    1490         \\

\hline
\end{tabular}
\end{center}
\end{table} 

Two sets of data have been collected with this device: a calibration using  external $^{232}$Th sources (65 h) and  a 
background measurement (66 h). The temperature of the detectors during our test  was $\sim$ 14 mK.
The intensity of the calibration sources was not optimized for small crystals, so that the statistics collected within the 
calibration peaks of ZnMoO$_4$ is rather  poor.
The energy resolution evaluated at 
911, 2615   and at 5407 keV ($^{228}$Ac, $^{208}$Tl and internal contamination of $^{210}$Po, respectively) are  
3.6$\pm$1, 5.6$\pm$2 and 6$\pm$1 keV, respectively.
The FWHM energy resolution of the LD, evaluated on the X doublet at  5.90 and 6.49  keV, is 470$\pm$20 eV.

\begin{figure}
\begin{center}
\resizebox{0.7\textwidth}{!}
{\includegraphics{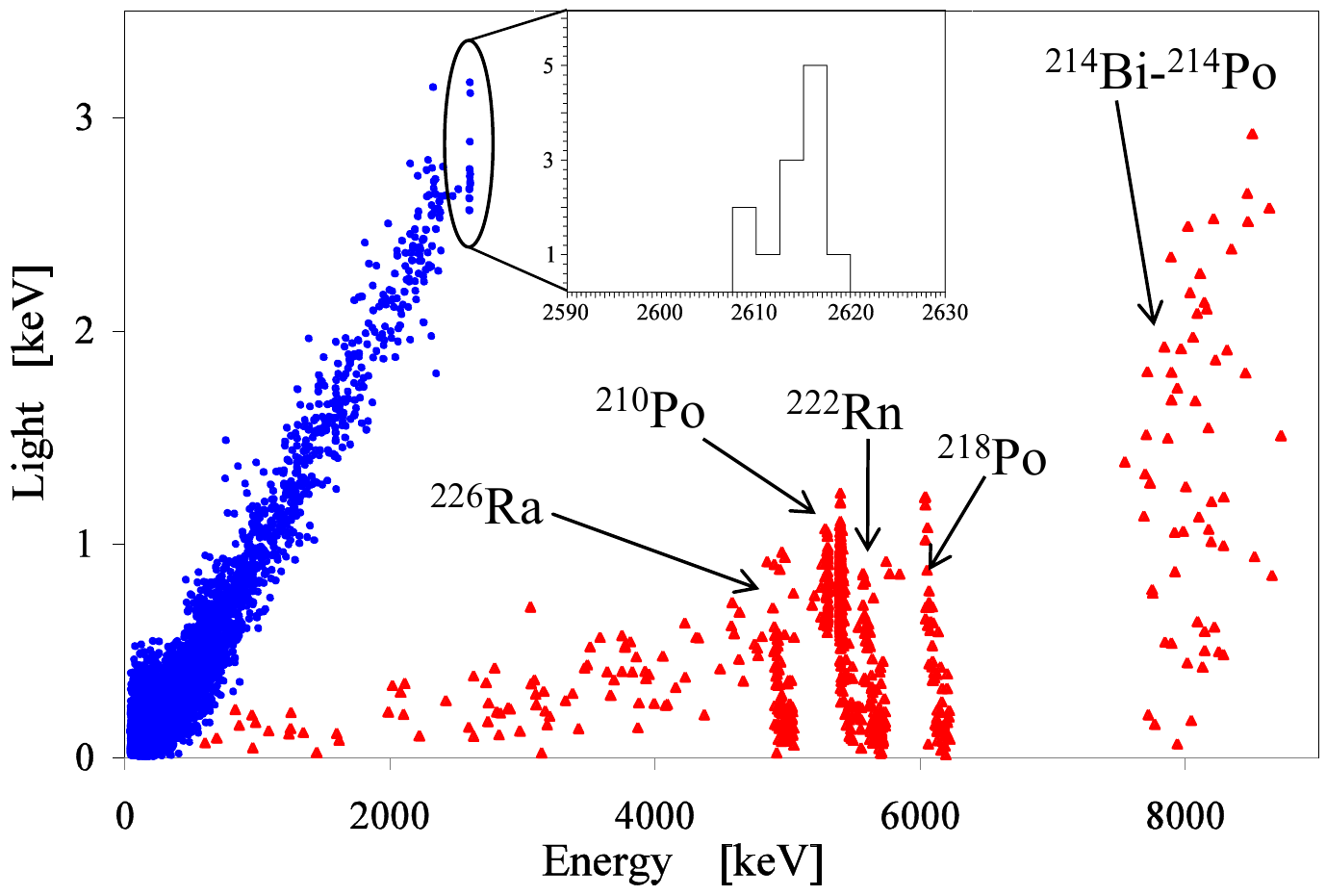}}
\caption{Scatter plot light vs. heat obtained with a $^{232}$Th calibration (65 h) plus a background measurement (66 h). The main observed lines 
are pointed out. In the inset the 2615 keV $\gamma$-line of $^{208}$Tl. The dark (blue) points are identified as $\gamma/\beta$ events, while the 
light (red) triangles are identified as $\alpha$-particles. The BiPo points are due to mixed events induced by  a $\gamma/\beta$ decay 
``immediately'' followed by an $\alpha$ decay, with $\tau$=164~$\mu$s, too fast for both our detectors to be distinguished from the parent $\gamma/\beta$.}
\label{fig:fig5}      
\end{center}
\end{figure}

In this field the usual way to present the results is to draw the light vs. heat scatter plot \cite{AP-2010}. 
Here each event is identified by a point with abscissa equal to the heat signal (recorded by the ZnMoO$_4$ bolometer), and 
ordinate equal to the light signal (simultaneously recorded by the LD). In the scatter plot, $\gamma/\beta$ and $\alpha$  give 
rise to  separate bands, in virtue of their characteristic light to heat ratio.
In Fig.~\ref{fig:fig5} we present the scatter plot light vs. heat of the sum of the two above mentioned measurements (Calibration + Background).
 
The LY of this crystal (measured on the 2615 keV $^{208}$Tl  line) can be  evaluated as 1.1 keV/MeV. This value is rather small. Using the same LD we evaluated the LY of a 
large (510 g)  CdWO$_4$ crystal to be 17.4 keV/MeV  \cite{AP-2010}. Nonetheless it is  evident from Fig.~\ref{fig:fig5} that the continuous  
background induced by $\alpha$-emitters located close to the surfaces is very efficiently recognized with respect to the $\gamma/\beta$ events.
The $\gamma/\beta$ band is mainly due to the interaction of $\gamma$ rays produced from
the calibration source or from radioactive contamination of the detector and its
set-up. The events in the $\alpha$ band are due to intrinsic radioactive  contamination of $^{226}$Ra, of the $^{238}$U chain . The
continuum below the  $^{226}$Ra peak can be ascribed to an alpha contamination in the
surfaces of the mounting structure, producing degraded alpha particles with a continuous distribution of energies. $^{214}$Bi-$^{214}$Po 
events are due to the fast decay chain of these two isotopes that produce a beta followed by an alpha, with a time separation below 
the integration time of the detector (i.e. they are recorded as a single event with an energy corresponding to the sum of $\alpha$ and
$\beta$ energies).

A strange feature has however to be pointed out: all the $\alpha$-lines  are characterized by 
a tail that draws negative slopes ``bands'', as it is apparent in  Fig.~\ref{fig:fig5}.
This effect is extremely evident in the $\alpha$-peaks induced 
by   $^{226}$Ra, $^{222}$Rn and $^{218}$Po.
This results in a large smearing of the energy resolution of these peaks. The  ``doublet'' due to $^{210}$Po, instead, shows this 
unexpected feature only -very marginally- on the full energy peak, at 5407 keV line.

This strange behavior  is likely when  the contaminations are close to the surfaces, both of the crystal and of the  the surrounding  materials.
The presence of large  quantity of surface contaminants, in fact, is corroborated by the large $\alpha$-continuum background, visible in the same figure. 
This issue should be studied in more details. This is however beyond the scope of this work. Moreover this effect -even if not well understood - increases 
the separation  between $\alpha$ and $\gamma/\beta$ interactions in the scatter plot.
Nonetheless, by assuming, in  a very  conservative way, that all the observed $\alpha$ lines are induced by internal contamination, we obtained the values 
in Tab.~\ref{Tab:tab3}.

\nopagebreak
\begin{table}[htb]
\caption{Radioactive $\alpha$ activity of ZnMoO$_4$ crystal. The activity  in $^{232}$Th corresponds to a limit in the  contamination  of 
7$\cdot 10^{-11}$  g/g. The contamination in  $^{238}$U has a limit of 2$\cdot 10^{-11}$  g/g. The observed $^{226}$Ra
contamination, instead, would correspond - assuming secular equilibrium- to a contamination  of $^{238}$U = 6$\cdot 10^{-10}$  g/g.}
\label{Tab:tab3}
\begin{center}
\begin{tabular}{|l|l|l|}
 \hline Chain                   & Nuclide            & Activity (mBq/kg) \\

 \hline
 $^{232}$Th              & $^{232}$Th         &   $\le$ 0.3             \\
  ~                      & $^{228}$Th         &   $\le$  0.3            \\
 \hline
 $^{238}$U               & $^{238}$U          & $ \le$ 0.2              \\
 ~     				     & $^{234}$U          & $ \le$ 0.8              \\
 ~                       & $^{230}$Th         & $ \le$ 0.3              \\
 ~                       & $^{226}$Ra         & 8.1 $\pm$0.3            \\
 ~                       & $^{210}$Po         &  28 $\pm$2              \\
 \hline
 Total $\alpha$ activity & ~                  & 73 $\pm$2               \\
 \hline

\end{tabular}
\end{center}
\end{table}

It can be seen that, as it very often happens, the $^{238}$U chain is broken at $^{226}$Ra. With respect to the  $^{232}$Th decay chain, 
the most dangerous \cite{PHAN-2006} internal contaminant  for this kind of detectors, we only observe a limit.

A very interesting feature of this compound comes from the observation that $\alpha$ and $\gamma/\beta$ events give rise to 
slightly different thermal pulses. In fact this promising feature was very recently exploited in CaMoO$_4$ \cite{Gironi-2010} as well as,
even if less  pronounced, in ZnSe \cite{ZnSe-2010}.

\begin{figure}
\begin{center}
\resizebox{1.0 \textwidth}{!}
{\includegraphics{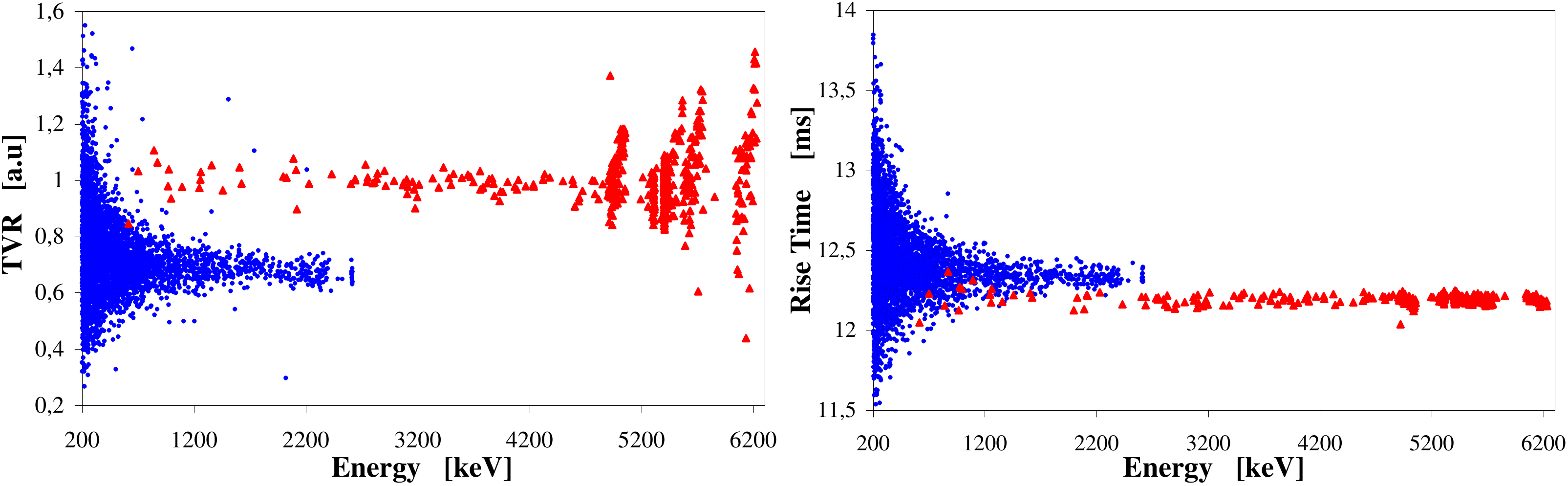}}
\caption{Pulse Shape Analysis of the events of Fig.\protect \ref{fig:fig5}. Left panel: TVR of the heat signal  in ZnMoO$_4$ vs. energy (heat).
The $\alpha$ events are effectively discriminated with respect to  $\gamma/\beta$ events down to low energies. Right panel: $\tau_{rise}$
of heat pulses vs. energy (heat) for the same events. It has to be pointed 
out, however, that the two parameters ($\tau_{rise}$ and TVR) are largely uncorrelated since one is calculated on the left part of the 
pulse maximum, while the latter on the right. Thus their combination will  further increase the discrimination factor.}
\label{fig:fig6}       
\end{center}
\end{figure}

In Fig.~\ref{fig:fig6} we show the Pulse Shape Analysis  performed only on the heat signal read in the ZnMoO$_4$ crystal.
The separation between $\alpha$ and $\gamma/\beta$ events is rather impressive, especially in the left panel of Fig.~\ref{fig:fig6}.
A comparison of the two distributions in the 2400--2620 keV energy region shows a separation of 7.2 $\sigma$.
This result needs further investigation with a high statistics comparison in the 3 MeV energy region (the $Q_{2\beta}$ value of $^{100}$Mo). 
It expresses however in a quantitative way the excellent rejection power of this technique.

\section{Conclusions}

Zinc molybdate (ZnMoO$_4$) single crystals were grown for the first time by the Czochralski method. 
Luminescence was measured under X ray excitation in the temperature range 85--400 K.
Emission maximum of ZnMoO$_4$ occurs at 544 nm at 295 K, and then shifts to 588 nm at 85 K. 

Applicability of ZnMoO$_4$ crystals as cryogenic phonon-scintillation detectors was demonstrated for the first time.
Despite the non  ``non perfect'' quality of our crystal sample, we obtained very encouraging results.   
The light output of ZnMoO$_4$ was estimated $<$ 6\% comparatively to CdWO$_4$. Even if rather small, the scintillation
signal permits a very efficient discrimination between  $\alpha$ and $\gamma$/$\beta$ particles.

Especially  this compound shows a different heat pulse shape between $\alpha$ and $\gamma$/$\beta$ events. This permits
the discrimination of the $\alpha$ induced background -\emph{without}  using the scintillation signal- at a level of 7.2 $\sigma$.
This feature could play a crucial role for a possible large scale double beta decay experiment.

\section{Acknowledgments}
The results reported here have been obtained in the framework of the Bolux R\&D Experiment funded by INFN, aiming  
at the optimization of a cryogenic DBD Experiment for a next generation experiment.
Thanks are due to E. Tatananni, A. Rotilio, A. Corsi and B. Romualdi  for continuous and constructive help in the  overall 
setup construction. Finally, we are especially grateful to Maurizio Perego for his invaluable help in the development and improvement  of the 
Data Acquisition software. 
The work of F.A.~Danevich and V.I.~Tretyak was partially supported by the Project ``Kosm\-omik\-rofi\-zyka-2'' (Astroparticle physics) of the
National Academy of Sciences of Ukraine.


\begin{thebibliography}{99}

 \bibitem{Moh07}  R.N.~Mohapatra, et al, \emph{Theory of neutrinos: a white paper}, 
 \href{http://dx.doi.org/10.1088/0034-4885/70/11/R02} {Rep. Prog. Phys.  \rm \bf 70 \rm (2007):1757} \href{http://arxiv.org/abs/hep-ph/0510213v2}{[hep-ph/0510213]}
 
 \bibitem{Fogli} G. L. Fogli, et al., \emph{Observables sensitive to absolute neutrino masses: A reappraisal after WMAP 3-year and first MINOS results},  
 \href{http://dx.doi.org/10.1103/PhysRevD.75.053001} {Phys. Rev. \rm \bf  D 75 \rm  (2007):053001} \href{http://arxiv.org/abs/hep-ph/0608060v1}{[hep-ph/0608060]}
 
 \bibitem{Fogli-2} G. L. Fogli, et al., \emph{Observables sensitive to absolute neutrino masses. II} 
 \href{http://dx.doi.org/10.1103/PhysRevD.78.033010}{Phys. Rev. \rm \bf  D 78 \rm  (2008):033010} \href{http://arxiv.org/abs/0805.2517v3}{[hep-ph:0805.2517]}
 
 \bibitem{Vissani} A. Strumia and F. Vissani, \emph{Neutrino masses and mixings and...}  (2010) 
 \href{http://arxiv.org/abs/hep-ph/0606054v3}{[hep-ph/0606054v3]}

 \bibitem{reviewElliot} S.R. Elliott, J. Engel, \emph{Double Beta Decay} 
 \href{http://dx.doi.org/10.1088/0954-3899/30/9/R01}{J. Phys. \rm \bf G 30 \rm (2004):183} 	\href{http://arxiv.org/abs/hep-ph/0405078v2}{[hep-ph/0405078v2]}

 \bibitem{Avi08}  F.T.~Avignone III, S.R. Elliott, J. Engel, \emph{Double beta decay, Majorana neutrinos, and neutrino mass},
 \href{http://dx.doi.org/10.1103/RevModPhys.80.481} {Rev. Mod. Phys.  \rm \bf 80 \rm (2008):481} 

 \bibitem{Sim-2008} F. Simkovic, et al., \emph{Anatomy of nuclear matrix elements for neutrinoless double-beta decay},
 \href{http://dx.doi.org/10.1103/PhysRevC.77.045503} {Phys. Rev. \rm \bf C 77 \rm (2008):045503} \href{http://arxiv.org/abs/0710.2055v3}{[nucl-th:0710.2055]}

 \bibitem{Civ-2009} O. Civitarese, J. Suhonen, \emph{Nuclear matrix elements for double beta decay in the QRPA approach: A critical review},
 \href{http://dx.doi.org/10.1088/1742-6596/173/1/012012} {J. Phys.: Conf. Ser. \rm \bf 173 \rm (2009):012012}

 \bibitem{Men-2009} J. Menendez, et al., \emph{Disassembling the Nuclear Matrix Elements of the Neutrinoless double beta Decay},
 \href{http://dx.doi.org/10.1016/j.nuclphysa.2008.12.005}  {Nucl. Phys. A \rm \bf 818 \rm (2009):139} \href{http://arxiv.org/abs/0801.3760v3}{[nucl-th:0801.3760]}

 \bibitem{BandI-2009} J. Barea and F. Iachello, \emph{Neutrinoless double-beta decay in the microscopic interacting boson model},
 \href{http://dx.doi.org/10.1103/PhysRevC.79.044301} {Phys. Rev. \rm \bf C 79 \rm (2009):044301}

 \bibitem{Aud03}  G.~Audi, A.H.~Wapstra, C.~Thibault, \emph{The AME 2003 atomic mass evaluation: (II). Tables, graphs and references}, 
 \href{http://dx.doi.org/10.1016/j.nuclphysa.2003.11.003} {Nucl. Phys.  \rm \bf A 729 \rm (2003):337}

 \bibitem{Boh05}  J.K.~Bohlke, et al., \emph{Isotopic Compositions of the Elements, 2001},
 \href{http://dx.doi.org/10.1063/1.1836764} {J. Phys. Chem. Ref. Data  \rm \bf 34 \rm (2005):57}

 \bibitem{Arn05}  R.~Arnold, et al., F\emph{irst Results of the Search for Neutrinoless Double-Beta Decay with the NEMO 3 Detector},
 \href{http://dx.doi.org/10.1103/PhysRevLett.95.182302} {Phys. Rev. Lett.  \rm \bf 95 \rm (2005):182302}

 \bibitem{Arna08} C.~Arnaboldi, et al., \emph{Results from a search for the 0-neutrino-decay of 130-Te},
 \href{http://dx.doi.org/10.1103/PhysRevC.78.035502} {Phys. Rev.  \rm \bf C 78 \rm (2008):035502}

 \bibitem{Gomez2007} H.~Gomez, et al., \emph{Background reduction and sensitivity for germanium double beta decay experiments},
 \href{http://dx.doi.org/10.1016/j.astropartphys.2007.08.008} {Astropart. Phys.  \rm \bf 28 \rm (2007):435} \href{http://arxiv.org/abs/0708.3987v1}{[nucl-ex:0708.3987]}

 \bibitem{Zde04}  Yu.G.~Zdesenko, F.A.~Danevich, V.I.~Tretyak, \emph{Sensitivity and discovery potential of the future 2beta decay experiments},
 \href{http://dx.doi.org/10.1088/0954-3899/30/9/002} {J. Phys. G: Nucl. Part. Phys.  \rm \bf 30 \rm (2004):971}

 \bibitem {Arnaboldi2004} C. Arnaboldi, et al., \emph{CUORE: a cryogenic underground observatory for rare events},
 \href{http://dx.doi.org/10.1016/j.nima.2003.07.067} {Nucl. Instr. and Meth. \rm \bf A 518 \rm(2004):775}  

 \bibitem {fondoBB} M. Pavan, et al., \emph{Control of bulk and surface radioactivity in bolometric searches for double-beta decay},
 \href{http://dx.doi.org/10.1140/epja/i2007-10577-0} {Eur. Phys. J \rm \bf A 36 \rm (2008):159}

 \bibitem{PHAN-2006} S. Pirro, et.al, \emph{Scintillating double-beta-decay bolometers},
 \href{http://dx.doi.org/10.1134/S1063778806120155} {Physics of Atomic Nuclei \rm \bf 69  \rm  (2006):2109} \href{http://arxiv.org/abs/nucl-ex/0510074v1} {[nucl-ex/0510074]}

 \bibitem{Belo05}  S.~Belogurov, et al.,  \emph{CaMoO4 Scintillation Crystal for the Search of 100-Mo Double Beta Decay},
 \href{http://dx.doi.org/10.1109/TNS.2005.852678} {IEEE Trans. Nucl. Sci. \rm \bf 52 \rm  (2005):1131}

 \bibitem{Ann08}   A.N. Annenkov, et al., \emph{Development of CaMoO4 crystal scintillators for a double beta decay experiment with 100-Mo},
 \href{http://dx.doi.org/10.1016/j.nima.2007.10.038} {Nucl. Instr. and Meth. \rm \bf A 584 \rm (2008):334}

 \bibitem{Mikh06a} V.B.~Mikhailik and H. Kraus, \emph{Cryogenic scintillators in searches for extremely rare events},
 \href{http://dx.doi.org/10.1088/0022-3727/39/6/026} {J. Phys. D: Appl. Phys. \rm \bf  39 \rm  (2006):1181}

 \bibitem{Dane10}  F.A.~Danevich, et al, \emph{Feasibility study of PbWO4 and PbMoO4 crystal scintillators for cryogenic rare events experiments},
 \href{http://dx.doi.org/10.1016/j.nima.2010.07.060}  {Nucl. Instr. and Meth. \rm \bf A 622 \rm (2010):608}

 \bibitem{Min92}   M. Minowa, et al., \emph{Measurement of the property of cooled lead molybdate as a scintillator},
 \href{http://dx.doi.org/10.1016/0168-9002(92)90945-Z} {Nucl. Instr. and Meth. \rm \bf A 320 \rm  (1992):500}

 \bibitem{Bash09}  N.V.~Bashmakova, {\it et al.}, Functional Materials \rm \bf 16 \rm (2009):266

 \bibitem{Bari09a} O.P.~Barinova, et al, \emph{Intrinsic radiopurity of a Li2MoO4 crystal} 
 \href{http://dx.doi.org/10.1016/j.nima.2009.06.003} {Nucl. Instr. and Meth. \rm \bf A 607 \rm (2009):573}
 
 \bibitem{Bari09b} O.P.~Barinova,et al, \emph{First test of Li2MoO4 crystal as a cryogenic scintillating bolometer},
 \href{http://dx.doi.org/10.1016/j.nima.2009.11.059} {Nucl. Instr. and Meth. \rm \bf A 613 \rm (2010):54}
 
 \bibitem{Bell07} P.~Belli, et al., \emph{Investigation of beta decay of 113Cd},
 \href{http://dx.doi.org/10.1103/PhysRevC.76.064603} {Phys. Rev. \rm \bf C 76 \rm (2007):064603}
 
 \bibitem{Ales98-b} A.~Alessandrello, et al., \emph{Measurements of internal radioactive contamination in samples of Roman lead to be used in experiments on rare events},
 \href{http://dx.doi.org/10.1016/S0168-583X(98)00279-1}  {Nucl. Instr. and Meth. \rm \bf B 142 \rm (1998):163}
 
 \bibitem{Dane09} F.A.~Danevich, et al., \emph{Ancient Greek lead findings in Ukraine},
 \href{http://dx.doi.org/10.1016/j.nima.2009.02.018}  {Nucl. Instr. and Meth. \rm \bf A 603 \rm (2009):328}
 
 \bibitem{Ivle08} L.I.~Ivleva, et al., \emph{Growth and properties of ZnMoO4 single crystals},
 \href{http://dx.doi.org/10.1134/S1063774508060266} {Crystallography Reports \rm \bf 53 \rm (2008):1087}
 
 \bibitem{Nago08} L.L.~Nagornaya, et al., \emph{Tungstate and Molybdate Scintillators to Search for Dark Matter and Double Beta Decay}
 \href{http://dx.doi.org/10.1109/TNS.2009.2022268} {IEEE Trans. Nucl. Sci. \rm \bf 56 \rm(2009):2513}
 
 \bibitem{Reic00} W.~Reichelt, et al., \emph{Mischkristallbildung im System CuMoO4/ZnMoO4},
 \href{http://dx.doi.org/10.1002/1521-3749(200009)626:9<2020::AID-ZAAC2020>3.0.CO;2-K}  {Zeit. fur anorg. und allg. Chemie \rm \bf 626 \rm (2000):2020}
 
 \bibitem {NIMA-2006} S. Pirro, et al., D\emph{evelopment of bolometric light detectors for double beta decay searches},
 \href{http://dx.doi.org/10.1016/j.nima.2005.12.009}   {Nucl. Instr. and Meth. \rm \bf A 559 \rm(2006):361}
 
 \bibitem {PRC-2008} C. Arnaboldi, et al., \emph{Results from a search for the 0 neutrino double beta decay of 130Te},
 \href{http://dx.doi.org/10.1103/PhysRevC.78.035502} {Phys. Rev. \rm \bf C 78 \rm (2008):035502}
 
 \bibitem {ALES98} A. Alessandrello, et al., \emph{Methods for response stabilization in bolometers for rare decays},
 \href{http://dx.doi.org/10.1016/S0168-9002(98)00458-6}  {Nucl. Instr. and Meth. \rm \bf A 412 \rm(1998):454}
 
 \bibitem {Arna2003} C. Arnaboldi, G. Pessina, E. Previtali, \emph{A programmable calibrating pulse generator with multi-outputs and very high stability},
 \href{http://dx.doi.org/10.1109/TNS.2003.815346} {IEEE Tran.  Nucl. Sci.  \rm \bf 50 \rm(2003):979}
 
 \bibitem {NIMA-2006-B} S. Pirro, \emph{Further developments in mechanical decoupling of large thermal detectors},
 \href{http://dx.doi.org/10.1016/j.nima.2005.12.197} {Nucl. Instr. and Meth. \rm \bf A 559 \rm(2006):672}
 
 \bibitem {NIMA-2006-C} C. Arnaboldi, G. Pessina, S. Pirro, \emph{The cold preamplifier set-up of CUORICINO: Towards 1000 channels},
 \href{http://dx.doi.org/10.1016/j.nima.2005.12.210} {Nucl. Instr. and Meth. \rm \bf A 559 \rm(2006):826}
 
 \bibitem {NIMA-2004} C. Arnaboldi, et al., \emph{The front-end readout for CUORICINO, an array of macro-bolometers and MIBETA, an array of micro-bolometers},
 \href{http://dx.doi.org/10.1016/j.nima.2003.11.319}  {Nucl. Instr. and Meth. \rm \bf A 520 \rm(2004):578}
 
 \bibitem{Arnaboldi-2004} C. Arnaboldi, et al., \emph{CUORE: a cryogenic underground observatory for rare events},
 \href{http://dx.doi.org/10.1016/j.nima.2003.07.067} {Nucl. Instrum. Meth.  \rm \bf A 518 \rm (2004):775}
 
 \bibitem{GattiManfredi-86} E. Gatti and P.F. Manfredi, Rivista del Nuovo Cimento \rm \bf 9 \rm (1986):1
 
 \bibitem{AP-2010} C. Arnaboldi, et al., \emph{CdWO4 scintillating bolometer for Double Beta Decay: Light and heat anticorrelation, light yield and quenching factors},
 \href{http://dx.doi.org/10.1016/j.astropartphys.2010.06.009} {Astrop. Phys. \rm \bf 34 \rm (2010):143} \href{http://arxiv.org/abs/1005.1239v1} {[nucl-ex:1005.1239]}
 
 \bibitem{Gironi-2010} L. Gironi, \emph{Scintillating bolometers for Double Beta Decay search},
 \href{http://dx.doi.org/10.1016/j.nima.2009.10.080}  {Nucl. Instr. and Meth. \rm \bf A 617 \rm (2010):478}
 
 \bibitem{ZnSe-2010} C. Arnaboldi, et al., \emph{Characterization of ZnSe scintillating bolometers for Double Beta Decay},
 \href{http://dx.doi.org/10.1016/j.astropartphys.2010.09.004} {accepted by Astropart. Phys.} \href{http://arxiv.org/abs/1006.2721v2} {[nucl-ex:1006.2721]}
 
\end{thebibliography}
\end{document}